\documentclass[amsmath,amssymb,twocolumn]{revtex4}
\usepackage{amssymb}
\usepackage{graphicx}
\usepackage[colorlinks=true,linkcolor=blue,citecolor=blue]{hyperref}

\begin{document}

\title{The system of correlation kinetic equations and the generalized equivalent circuit for hopping transport}
\author{A.V. Shumilin$^1$, Y.M. Beltukov$^1$}

\affiliation{$^1$Ioffe Institute, 194021 St.-Petersburg, Russia}

\begin{abstract}
We derive the system of equations that allows to include non-equilibrium correlations of filling numbers into the theory of the hopping transport. The system includes the correlations of arbitrary order in a universal way and can be cut at any place relevant to a specific problem to achieve the balance between rigor and computation possibilities. In the linear-response approximation, it can be represented as an equivalent electric circuit
that generalizes the Miller-Abrahams resistor network. With our approach, we show that non-equilibrium correlations are essential to calculate conductivity and distribution of currents in certain disordered systems. Different types of disorder affect the correlations in different applied fields. The effect of energy disorder is most important at weak electric fields while the position disorder by itself leads to non-zero correlations only in strong fields.
\end{abstract}

\maketitle

\section{Introduction}

Perhaps the most known tool for description of hopping transport is the Miller-Abrahams resistor network \cite{MA}. It reduces the problem of hopping transport in small electric fields to the equivalent electric circuit. Its nodes, which represent the localization sites, are connected by resistors. The conductivity of the circuit can be then analyzed within the framework of the percolation theory \cite{Efr-Sh} or calculated numerically as a solution of Kirchhoff equations.

The Miller-Abrahams network is based on the mean-field approximation. In this approximation the averaged product of two filling numbers $\overline{n_i n_j}$ is considered to be equal to the product of averaged filling numbers
$\overline{n}_i \overline{n}_j$. Therefore, all the correlations of site filling numbers are neglected.  The simplest neglected correlation can be expressed as the covariation of two filling numbers $\overline{c}_{ij} = \overline{n_i n_j} - \overline{n}_i \overline{n}_j$. The mean field approximation is uncontrollable as discussed in \cite{Efr-Sh}.

The most obvious reason for the presence of the correlations $\overline{c}_{ij}$ is the Coulomb interaction between electrons on sites $i$ and $j$. It is known that when this interaction dominates the energy disorder, the system is strongly correlated and shows a number of glassy features such as a number of pseudo-ground states and slow relaxations \cite{Col-glass-1,Col-glass-2,Kogan}. We would like to note that even in this case the mean field approximation is widely applied. When the density of states with the Coulomb gap is considered and self-interaction is excluded from the energy of phonon participating in the hop, the mean field approximation leads to correct temperature dependence of conductivity, the Efros-Shlkovskii law. However, even if the Coulomb interaction is neglected the correlations can appear due to non-equilibrium response to applied electric field. The present study is focused on these non-equilibrium correlations and therefore we do not consider the Coulomb interaction between electrons on different sites and the corresponding physical phenomena. However, the on-site Coulomb repulsion is considered to be strong preventing double occupation.

The contribution of non-equilibrium correlations to the transport properties was first discussed by P.M. Richards \cite{Richards}.
In Ref.~\cite{Richards} the chain of two kinds of sites was studied with Monte-Carlo simulation. The conductivity of the chain was found to be substantially different from the result of Miller-Abrahams model due to the contribution of non-equilibrium correlations. The importance of correlations was confirmed by Chase and Thouless \cite{Thouless}. To take the correlations into account an analog of Miller-Abrahams network was proposed.  A node of network in Ref.~\cite{Thouless} represents a many-electron configuration. This method includes all the possible correlations but the apparent network contains $2^N$ nodes where $N$ is the number of sites in the system. It allows to apply the method to very small systems only. The detailed study of correlations in one-dimensional chain (both periodic and random) was made in \cite{Pitis91, Pitis92, PitisMonte93} where the effect of correlations on conductivity and diffusion was estimated both analytically and with Monte-Carlo simulations.

Different opinions exist concerning the importance of non-equilibrium correlations in strongly disordered systems being in the Mott law regime. Levin, Nguen, and Shklovskii \cite{Shklovskii-cor} calculated the conductivity of a cubic system with 1000 sites with Monte-Carlo simulation that automatically takes into account all the non-equilibrium correlations. The results of simulation were compared with numerical solution of Kirchhoff equations for Miller-Abrahams resistor network. The difference of the conductivities calculated with the two methods was of the order of unity and was considered to be small compared to the exponential dependence of current on temperature and localization radius. The results of the simulation were explained with 4-site model that was solved analytically. On the other hand, recently it was pointed out \cite{Aleiner1} that correlations can substantially decrease the conductivity of resistor between hoping sites compared to the prediction of Miller-Abrahams model. It occurs for the sites with different sign of $\varepsilon_i - \mu$ where $\varepsilon_i$ is the site energy and $\mu$ is the chemical potential.  The authors of \cite{Aleiner1} concluded that the current should form two separate percolation clusters with quite rare connections. Two color percolation model was proposed to describe the conductivity. The approach applied in \cite{Aleiner1} allows to consider pair correlations in close pairs of sites (and only them) and treat them as a modification of Miller-Abrahams resistors.

Another reason to study the non-equilibrium correlations appeared recently due to the development of organic semiconductors where transport usually occurs due to polaron hopping. The correlations of site filling numbers in organic materials were discussed in \cite{pol-cor, AVS-cor}. In particular, it was understood that so-called organic magnetoresistance (OMAR) \cite{OMAR, OMAR2, OMAR3, Bobbert} observed in many organic semiconductors with hopping transport is
absent in the mean field framework \cite{AVS-MF}. This phenomenon is closely related to intersite correlations \cite{Bobbert-Res, dis-org, AVS-cor}. It reappears in the conventional theory when at least pair correlations in close sites are taken into account \cite{AVS-cor}. The similar mechanism of magnetoresistance was also proposed for ordinary semiconductors \cite{Aleiner2}
but to the best of the author knowledge was never observed in non-organic materials.

However, the further development of the theoretical aspects concerning the effect of the correlations on transport is hindered by the absence of a general framework that allows the theoretical description of these correlations. Only the pair correlations in close pairs can be included in Miller-Abrahams network as a modification of its resistors. However, this approximation is also uncontrollable and cannot be used to describe all the correlation-related phenomena. E.g., it cannot be used to rigorously describe the effect of spin relaxation on the conductivity when the spin relaxation is due to a mechanism that is closely related to hops. The examples of such mechanisms are the hyperfine interaction with atomic nuclei and spin-orbit interaction. This interplay between the spin relaxation and the current flow is the essence of the organic magnetoresistance and should be described in details before the quantitative theory of OMAR can be developed. It can be concluded that a new theoretical framework should be developed allowing to effectively include arbitrary correlations into the theory.

The main goal of the present study is to develop a new method to take into account the non-equilibrium correlations. The method is based on the  Bogoliubov-Born-Green-Kirkwood-Yvon chain of equations \cite{Bogolubov,Landau10,Balescu} and generalizes the mean-field approximation to take into account the correlations of arbitrary order. We derive the system of correlation kinetic equations (CKE) that describes all possible correlations of site filling numbers in a universal way and relate them to the currents between sites.  When the applied electric field
is weak CKE can be reduced to an effective electric circuit that generalizes the Miller-Abrahams resistor network. The nodes of the circuit in our approach represent the localization sites and all the possible correlations of site filling numbers.  When all the correlations are included  our method is as exact as the resistor network \cite{Thouless} and also includes $2^N$ nodes. However, we show that correlations between  distant sites and high order correlations are small. It allows one to cut the circuit at some point and consider a number of nodes equal to $const \times N$. The simplest possible cutoff leads to the Miller-Abrahams network. The next-to-simplest cutoff leads the approximation exploited in Ref. \cite{Aleiner1}.
In the present study we will focus on the simplest situation where Coulomb interaction between distant sites and electron spin are not included into the theory.

The article is organized as follows. In Sec. \ref{sect-kin} we develop the system of equations that describes the correlations. In particular, in Sec. \ref{sect-kin-lin} we derive it for the small applied electric field and represent as an equivalent electric circuit. In Sec. \ref{sect-kin-nl} we derive a more general and complex system of equations that should be used far from equilibrium. In Sec. \ref{sect-cur} we use the developed method to study the effect of correlations on the currents in different disordered systems. In particular, in Sec. \ref{sect-cur-mott} we discuss the effect of correlations on conductivity and current distribution in semiconductors in the Mott law regime. In Sec. \ref{sect-cur-nl} we describe the effect of correlations in strong electric fields and discuss the difference between position and energy disorder. In Sec. \ref{sect-monte} we compare our method with Monte-Carlo simulation. In Sec. \ref{sect-dis} we present the overall discussion of the obtained results.

\section{Correlation kinetic equations}
\label{sect-kin}

In this section we give a general framework for the description of the non-equilibrium correlations. We derive the kinetic equations that relate the correlations to one another and to the applied electric field. We do not consider long-range Coulomb interaction between different sites. However, the on-site Coulomb interaction is strong preventing the double occupation of sites.

In the equilibrium the averaged filling numbers are equal to the Fermi function
\begin{equation}\label{n0}
\overline{n}_i = n_i^{(0)} = \frac{1}{\exp\left( \frac{\varepsilon_i}{T}\right) + 1}.
\end{equation}
Here index $i$ enumerates the hopping site. $\varepsilon_i$ is the site energy related to the Fermi level. $n_i$ is its filling number that can be equal to either $1$ or $0$. Line over $n_i$ indicates time or ensemble averaging.

In the equilibrium the filling numbers are independent. It means that
\begin{equation}
\overline{n_i n_j} = \overline{n}_i\overline{n}_j = n_i^{(0)}n_j^{(0)}.
\end{equation}

When the system is driven out of equilibrium, e.g. due to an application of an external electric field, the average filling numbers can be different from the equilibrium ones. We define $\overline{\delta n_i} = \overline{n_i - n_i^{(0)}}$. The kinetics of $\overline{\delta n_i}$ is governed by charge conservation law
\begin{equation} \label{kin_n}
\frac{d}{dt}\overline{\delta n_i} = \sum_j J_{ij}.
\end{equation}
Here $J_{ij}$ is the electron flow from site $j$ to site $i$. It can be expressed as
\begin{multline}\label{cur}
J_{ij} = W_{ij} (1-\overline{n}_i) \overline{n}_j - W_{ji}(1-\overline{n}_j) \overline{n}_i  + \\
(W_{ji} - W_{ij})(\overline{n_i n_j} - \overline{n}_i \overline{n}_j).
\end{multline}
Here $W_{ij}$ is the rate of hopping from site $j$ to site $i$ with emission or absorption of a phonon. In our study we adopt the simplified expression for $W_{ij}$ for definiteness:
\begin{equation} \label{Wij}
W_{ij} = \gamma_0 |t_{ij}|^2 \times \min \left[1, \exp\left( \frac{\varepsilon_j - \varepsilon_i}{T}\right) \right].
\end{equation}
Here $\gamma_0$ is the constant describing the electron-phonon interaction, $t_{ij}$ is the overlap integral between the sites $i$ and $j$ and $T$ is temperature. However, the details of expression (\ref{Wij}) are not particularly important for the study provided that the detailed balance holds in the equilibrium ${W_{ij}(1-n_i^{(0)})n_j^{(0)} = W_{ji}(1-n_j^{(0)}) n_i^{(0)}}$.
The last term in (\ref{cur}) describes the effect of correlations on the currents. It is neglected in the mean field approximation
and in the Miller-Abrahams resistor network.

Rate equations similar to (\ref{kin_n}) can be written for pair correlations. However, these equations will contain correlations of higher orders. If all the correlations are taken into account the apparent chain of equations will be the exact description of system kinetics. The number of correlations is extremely large. It is equal to $2^N$ where $N$ is the number of sites. However, it is natural to assume that correlations of filling numbers on distant sites are small and can be neglected. It leads to the possibility to cut the chain.

The idea of the present study is to write the chain of equations in a general form that allows to place the cutoff at any place relevant for a given problem. We derive the equations in two situations. In the first one, the applied electric field and all the correlations are assumed to be small. In the second situation, we consider an arbitrary non-equilibrium state of the system that leads to more cumbersome equations.

\subsection{Linearized equations}
\label{sect-kin-lin}

In this section we consider the system that is close to equilibrium. It is drawn from the equilibrium due to the small applied electric field described by on-site electric potentials $\varphi_i = {\bf r}_i \cdot {\bf E}$. We keep only first order perturbations proportional to $\varphi_i$.

The perturbations of averaged filling numbers $\overline{\delta n}_i$ are proportional to applied field. To describe them we  linearize equations (\ref{kin_n}),(\ref{cur}).
\begin{equation}\label{kinlin1}
\frac{d}{dt} \overline{\delta n}_i = \sum_j T_{ij} \overline{\delta n}_j - T_{ji}\overline{\delta n}_i + D_{ij}\overline{\delta n_i \delta n_j} + S_{ij}.
\end{equation}
Here $T_{ij} = W_{ij}^{(0)}(1-n_i^{(0)}) + W_{ji}^{(0)} n_i^{(0)}$ is the rate of transition of additional charges from site $j$ to site $i$. The index $(0)$ in $W_{ij}^{(0)}$ indicates that we consider equilibrium values of hopping rates. $S_{ij}$
describes the currents generated directly by the applied electric potentials $\varphi_i$
\begin{equation}
S_{ij} = W_{ij}^{(0)}(1-n_i^{(0)})n_j^{(0)} \frac{e \varphi_j - e\varphi_i}{T}.
\end{equation}

The term $D_{ij}\overline{\delta n_i \delta n_j}$ describes the contribution of the non-equilibrium correlations to the currents. Although the products of averaged corrections to the filling numbers  $\overline{\delta n}_i \overline{\delta n}_j$ are small $\propto \varphi_i^2$ and should be neglected, we keep the averaged product $\overline{\delta n_i \delta n_j}$. In the linear approximation, it is equal to covariation of filling numbers on sites $i$ and $j$.
\begin{equation}\label{lcov2}
\overline{\delta n_i \delta n_j} = \overline{(n_i - n_i^{(0)})(n_j - n_j^{(0)})} \approx \overline{n_i n_j} - \overline{n}_i \overline{n}_j.
\end{equation}
Note that the values $(n_i - n_i^{(0)})(n_j - n_j^{(0)})$ are not small without the averaging even in the equilibrium. $n_i$ and $n_j$ are equal to 0 or 1. Therefore, the smallness of covariation (\ref{lcov2}) is controlled not by the smallness of $\delta n_i = n_i - n_i^{(0)} \sim 1$ but by the close-to-equilibrium statistics of filling numbers.

The averaged product $\overline{\delta n_i \delta n_j}$ describes the pair correlation between sites $i$ and $j$. However, we would like to consider the correlations of arbitrary order. Therefore, we introduce notations that can be used for any number of the involved sites. Let $I$ be some set of sites $I = \{ i_1,i_2,...,i_{K} \}$ where $K$ is the number of sites in the set. We denote
\begin{equation}
\overline{\delta n}_I = \overline{\delta n_{i_1}\cdot \delta n_{i_2} \cdot ... \cdot \delta n_{i_K} }.
\end{equation}
In the linear approximation in terms of applied electric field, $\overline{\delta n}_I$ is equal to the $K$-th order covariation of the filling numbers in the set $I$.

To derive the rate equation for $\overline{\delta n}_I$ we note that these covariations can be changed due to the hops involving at least one site from $I$. Different hops are considered to be independent. It means that the time derivative of $\overline{\delta n}_I$ can be expressed as follows:
\begin{equation}\label{ddt}
\frac{d}{dt} \overline{\delta n}_I = \sum_{i \in I, k \notin  I} \left( \frac{d}{dt} \right)_{ik} \overline{\delta n}_I +
\sum_{i,j \in I} \left( \frac{d}{dt} \right)_{ij} \overline{\delta n}_I.
\end{equation}
Here notation $(d/dt)_{ij}$ stands for the term in the expression of a time derivative of some physical quantity related to hops $i \leftrightarrow j$. Each pair of sites $ij$ is included into summation only once. We will show that the first term in (\ref{ddt}) corresponds to transitions between the correlation of the same order due to electron diffusion. The second term corresponds to the generation of higher-order correlations by the lower-order ones.

To give the explicit expression for the terms in (\ref{ddt}) let us consider a term related to some given sites $i$ and $k$ in the expression (\ref{kinlin1}). Any quantity $A$ independent of filing numbers on the sites $i$ and $k$ is not changed due to these hops.
It leads to the expression
 \begin{multline}\label{ddtA}
\left(\frac{d}{dt} \right)_{ik} \overline{\delta n_i A} = \\  T_{ik} \overline{\delta n_k A} - T_{ki}\overline{\delta n_k A} + D_{ik}\overline{\delta n_i \delta n_k A} + S_{ik} \overline{A}.
\end{multline}
Here $A$ is an arbitrary quantity that can depend on any filling numbers besides $n_i$  and $n_k$.

The general equation (\ref{ddtA}) allows derivation of the first term $(d/dt)_{ik}$ with the selection $A = \delta n_{I \backslash i}$. Here notation $I \backslash i$ stands for the set difference of sets $I$ and $\{i \}$.
\begin{multline}\label{dtl-ik}
\left(\frac{d}{dt}\right)_{ik} \overline{\delta n_{i \cup I}} = T_{ik} \overline{\delta n_{k \cup I}} - T_{ki}\overline{\delta n_{i \cup I}} + \\
D_{ik}\overline{\delta n_{i \cup k\cup I}} + S_{ik} \overline{\delta n_{I}}.
\end{multline}
Note that in most cases the term $S_{ik} \overline{\delta n_{I}}$ is of the second order in terms of applied electric field and should be neglected. The only exception is the case $I = \emptyset$, which describes transitions between corrections $\overline{\delta n}_i$ to occupation numbers. In this case $\overline{\delta n_{\emptyset}} = 1$ and the term $S_{ik} \overline{\delta n_{I}}$ describes the usual contribution of the electric potential to the currents.

To derive the second term in (\ref{ddt}) we note that the averaged product $\overline{n_in_j}$ cannot be changed due to hops $i \leftrightarrow j$. Naturally, $\overline{n_in_j}$ is the probability of joint occupation of sites $i$ and $j$, while $i \leftrightarrow j$ hops are impossible when both sites are occupied. It yields
\begin{multline}\label{ddtA2}
\left( \frac{d}{dt} \right)_{ij} \overline{\delta n_i \delta n_j A} =
 \\
 - n_j^{(0)}\left( \frac{d}{dt} \right)_{ij} \overline{\delta n_i  A} - n_i^{(0)}\left( \frac{d}{dt} \right)_{ij} \overline{\delta n_j  A}.
\end{multline}
Here similarly to Eq.~(\ref{ddtA}) we added an arbitrary quantity $A$ independent of filling numbers $n_i$ and $n_j$.

The expression (\ref{ddtA2}) with  $A = \delta n_{I \backslash i \backslash j}$ leads to the following expression for the second term in (\ref{ddt})
\begin{multline}\label{dtl-ij}
\left(\frac{d}{dt}\right)_{ij} \overline{\delta n_{i \cup j\cup I}} = \left(n_i^{(0)} - n_j^{(0)} \right) \times \\
\left( T_{ij} \overline{\delta n_{j\cup I}} - T_{ji} \overline{\delta n_{j\cup I}} + D_{ij} \overline{\delta n_{i \cup j\cup I}} + S_{ij} \overline{\delta n_{I}}\right).
\end{multline}

The expressions (\ref{ddt}), (\ref{dtl-ik}), (\ref{dtl-ij}) compose the closed system of correlation kinetic equations for the hopping transport. To cut the system at some point one should simply neglect some correlations and consider $\overline{\delta n}_I = 0$ when $I$ is the set of sites with correlation neglected.

\begin{figure}[htbp]
   \centering
       \includegraphics[width=\columnwidth]{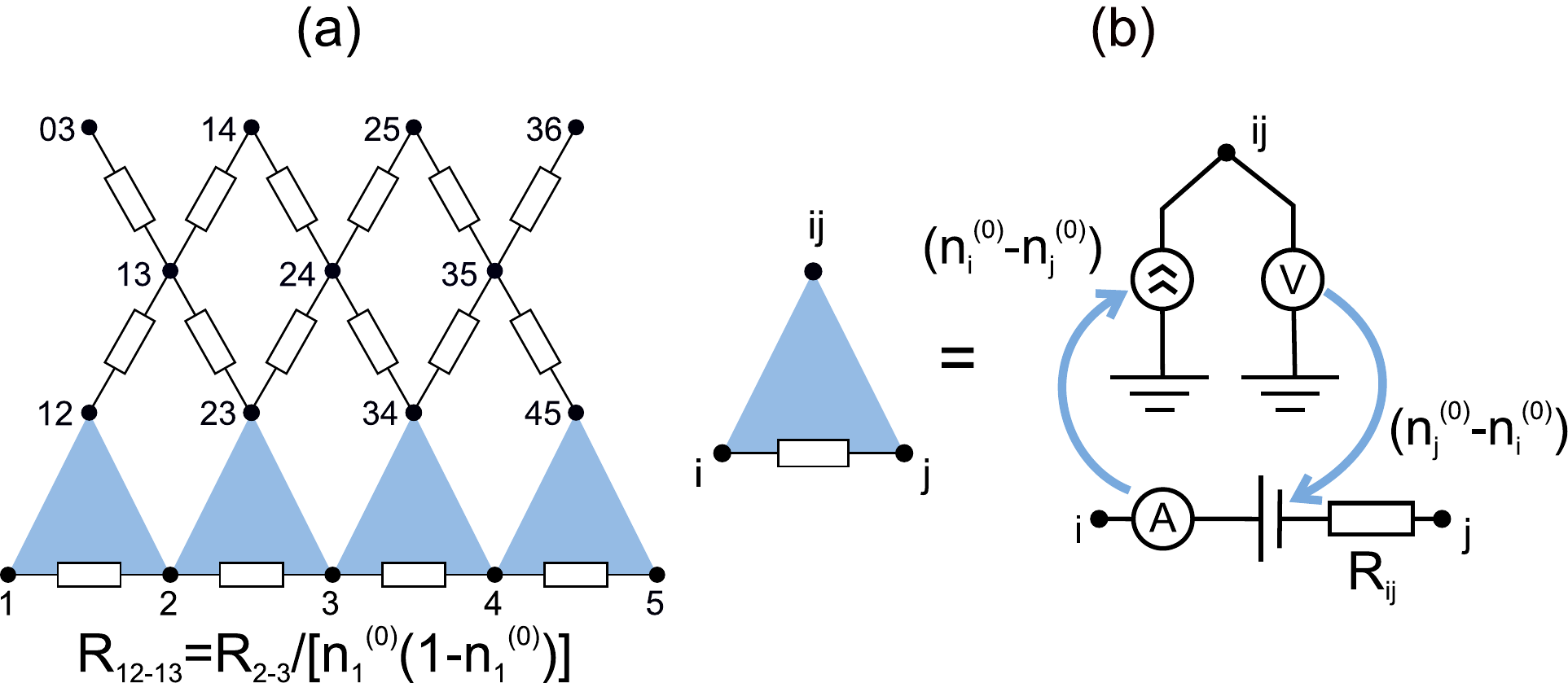}
       \caption{Part of equivalent circuit representing five sites and pair correlations between them (a). The connection between pair correlation layer and charge layer is shown with blue triangles. The triangles are explained in panel (b).}
   \label{fig1}
\end{figure}

The linearized system of CKE (\ref{ddt}), (\ref{dtl-ik}), (\ref{dtl-ij}) allows the simple representation with an equivalent electric circuit. To construct it let us consider the correlation in some set $I$ of sites that are far away from each another so that the direct hopping between sites within $I$ is impossible. In this case only the term $(d/dt)_{ik}$ in (\ref{ddt}) remains. One can see that $(d/dt)_{ik} \overline{\delta n_{I\cup i}} = - (d/dt)_{ik} \overline{\delta n_{I \cup k}}$. Therefore, when the second term in (\ref{ddt}) is neglected, the sum of all the covariations of any given order is conserved just like the sum of all occupation numbers. It allows to introduce the ``correlation currents''
\begin{equation}\label{gencur}
J_{ik}^{\{I\}} = \left( \frac{d}{dt} \right)_{ik} \overline{\delta n}_{i\cup I} = - \left( \frac{d}{dt} \right)_{ik} \overline{\delta n}_{k\cup I}.
\end{equation}
Here $I$ is any set of sites that does not include $i$ and $k$. When $I = \emptyset$ the expression (\ref{gencur}) gives the ordinary particle flow between sites $k$ and $i$.

It is also instructive to introduce the ``correlation potentials'' $\overline{u}_I$, related to $\overline{\delta n}_I$ as follows
\begin{equation}\label{udef}
\overline{\delta n}_I = \overline{u}_I \times \prod_{i \in I} n_i^{(0)}(1-n_i^{(0)}).
\end{equation}
When $I$ contains only one site $i$, the potential $\overline{u}_i$ is related to the current-induces correction $\delta\mu_i$ to the chemical potential on the site $i$ \cite{Efr-Sh}, $\overline{u}_i = \delta\mu_i/T$.  The expression for currents $J_{ik}^{\{I\}}$ in terms of potentials is
\begin{multline}\label{J1}
J_{ik}^{\{I\}} = \Gamma_{ik}^{\{I\}} \times
\\
\left[\overline{u}_{k\cup I} - \overline{u}_{i \cup I} + s_{ij}^{\{I\}} + \left(n_k^{(0)} - n_i^{(0)} \right) \overline{u}_{i \cup k \cup I}\right].
\end{multline}
Here $s_{ij}^{\{I\}}$ is the source term caused by the applied electric field: $ s_{ij}^{\{I\}} = (e/T)\,{\bf r}_{ij}\cdot {\bf E}$ if $I=\emptyset$ and zero otherwise.
One can consider correlations $\overline{\delta n}_{i\cup I}$ and $\overline{\delta n}_{k\cup I}$ to correspond to nodes of some electric circuit. $\Gamma_{ik}^{(I)}$ corresponds to the conductivity of resistor connecting these nodes.
\begin{equation}
\Gamma_{ik}^{\{I\}} = W_{ik}^{(0)}(1-n_i^{(0)}) n_k^{(0)} \prod_{l\in I} n_l^{(0)}(1-n_l^{(0)}).
\end{equation}
For any correlations $\Gamma_{ik}^{\{I\}} = \Gamma_{ki}^{\{I\}}$.

The hops inside the subset $I$ lead to additional current $\widetilde{J}_I$ that enters the node $\overline{\delta n}_I$. This current can be calculated as follows.
\begin{equation}\label{J2}
\widetilde{J}_I = \sum_{i,j \in I} \widetilde{J}_I^{(ij)}, \quad  \widetilde{J}_I^{(ij)} = \left( n_i^{(0)} - n_j^{(0)}\right) J_{ij}^{\{I \backslash i \backslash j\}}.
\end{equation}

Therefore the set of linearized CKE corresponds to an equivalent electric circuit. It is divided into levels corresponding to orders of correlations. Inside one level the circuit is a resistor network. The potentials in upper level generate addition voltage sources at lower level according to Eq.~(\ref{J1}). The currents in lower levels generate additional currents $\widetilde{J}_I$ that flow from ground to the upper levels according to Eq.~(\ref{J2}).
The part of this circuit representing a chain of five sites and pair correlations of filling numbers of these sites is shown in Fig.~\ref{fig1}.

The total current flow into any node of the circuit is equal to zero. It yields the system of Kirchhoff equations
\begin{equation}
\sum_{i \in I, \,k \notin I} J_{ik}^{\{ I \backslash i \}} + \sum_{i,j \in I} \widetilde{J}_I^{(ij)} = 0.
\label{eq:Kirchhoff}
\end{equation}
Each pair of sites $ij$ is included into summation only once similarly to Eq.~(\ref{ddt}). $I$ is any set of sites.  It corresponds to the node of the circuit representing the correlation inside this set.
If the currents are expressed in terms of potentials $\overline{u}_I$ the system (\ref{eq:Kirchhoff}) becomes the system of linear equations for these potentials.

Even in an ordinary resistor network, the solution of Kirchhoff equations has some degree of uncertainty. Any constant can be added to all the potentials without changing the currents and breaking the equations. In our case the uncertainty is even stronger. Each level of the circuit produces the transformation of potentials that keeps all the currents. The transformation related to the level $K$ can be described as follows. Arbitrary constant $c$ is added to all the potentials $\overline{u}_I$ for all sets $I$ with size $K$. At all the sets $I_k = \{i_1,i_2,...,i_{K-k} \}$ with size $K-k$ the potentials are also modified
\begin{equation}
\overline{u}_{I_k} \rightarrow \overline{u}_{I_k} + c\times (-1)^k \xi^{(k)}\left(n_{i_1}^{(0)}, ..., n_{i_{K-k}}^{(0)} \right).
\end{equation}
Here the notation $\xi^{(k)}(a,b,c,...)$ stands for the sum of all possible products of the values in the brackets with total power equal to $k$. For example, $\xi^{(3)}(a,b) = a^3 + a^2b + ab^2 + b^3$.

The uncertainty in potentials means that the Kirchhoff equations should be supplemented with boundary conditions to calculate the correlations in a real physical system. The most natural boundary condition state that correlation in any set $I$ should decrease to zero when the distance between sites of this set tends to infinity. This condition is automatically met when the system of kinetic equations is cut. In this case the only uncertainty is the arbitrary reference point of electrical potentials.

\subsection{Non-linear equations}
\label{sect-kin-nl}

When the system is far from equilibrium, the separation of averaged filling numbers $\overline{n}_i$ into equilibrium values $n_i^{(0)}$ and corrections $\overline{\delta n}_i$ is not instructive because the corrections are not small. The hopping rates $W_{ij}$ are substantially different from $W_{ij}^{(0)}$.  However, the description of the system with correlation kinetic equations is still possible. Also one can still hope that correlations between distant sites are small and the system of CKE can be cut.

The non-linear system of correlation kinetic equations relates the covariations of site filling numbers. We define the covariation $\overline{c}_I$ where $I$ is some set of sites
\begin{equation}
\overline{c}_I = \overline{\prod_{i\in I} (n_i - \overline{n}_i)}.
\end{equation}
Here $\overline{n}_i$ is the ensemble-averaged number of electrons on site $i$. By definition $\overline{c}_I = 0$ when $I$ contains only one site. Therefore, the variables of kinetic equations are the averaged numbers of sites $\overline{n}_i$ and the covariations $\overline{c}_I$ for sets $I$ containing at least two sites.

The separation of rate equations into the terms corresponding to different hops is still possible far from equilibrium:
\begin{equation}\label{nl-divis}
\frac{d}{dt}\overline{c}_I = \sum_{i\in I, k \notin I} \left( \frac{d}{dt}\right)_{ik} \overline{c}_I + \sum_{i,j \in  I}\left( \frac{d}{dt}\right)_{ij} \overline{c}_I.
\end{equation}
The general expression similar to Eq. (\ref{ddtA}) that allows the derivation of different terms for correlations reads
\begin{equation}\label{ddtAnl}
 \left( \frac{d}{dt}\right)_{ik} \overline{n_i A} = W_{ik}\overline{(1-n_i)n_k A} - W_{ki} \overline{(1-n_k)n_iA},
\end{equation}
where $A$ is independent of $n_i$ and $n_k$. The r.h.s. of Eq.~(\ref{ddtAnl}) should be decoupled into averaged values  and covariations.

With Eq.~(\ref{ddtAnl}) we derive
\begin{equation}\label{nl-JI-def}
\left( \frac{d}{dt}\right)_{ik} \overline{c}_{i\cup I} = - \left( \frac{d}{dt}\right)_{ik} \overline{c}_{k\cup I} = J_{ik}^{\{I\}},
\end{equation}
where we assumed that the set $I$ does not contain sites $i$ and $k$.  The expression for the correlation current is
\begin{multline}\label{nl-ik}
J_{ik}^{\{ I \}} =  \left(W_{ik} (1-\overline{n}_i) + W_{ki} \overline{n}_i \right) \overline{c}_{I \cup k} - \\
\left(W_{ki} (1-\overline{n}_k) + W_{ik} \overline{n}_k \right) \overline{c}_{I \cup i} + \\
(W_{ki} - W_{ik}) (\overline{c}_{i\cup k \cup I} - \overline{c}_{\{i,k\}}\overline{c}_I).
\end{multline}

\begin{figure*}[htbp]
   \centering
       \includegraphics[width=1.0\textwidth]{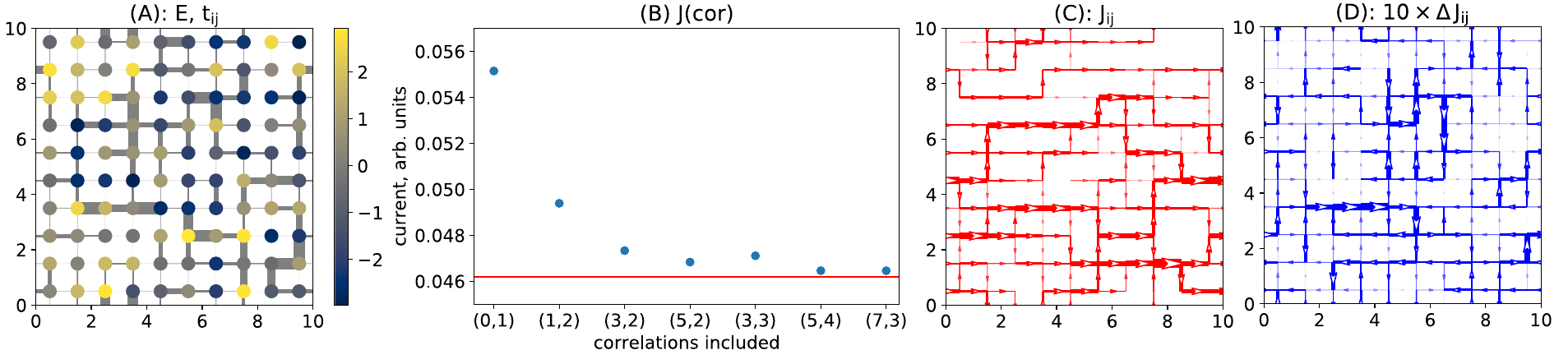}
       \caption{Results of the calculation in $10 \times 10$ lattice. (A) The structure of the system. (B) the dependence of total current on the correlations included into the computation. (C) the distribution of currents in the sample. (D) the difference in current distributions when correlations are neglected and are taken into account.}
   \label{fig:SmSys}
\end{figure*}

To derive the second term in (\ref{nl-divis}) we recall that probability of joint occupations of sites $i$ and $j$ cannot be changed due to $i \leftrightarrow j$ hops. Therefore, ${(d/dt)_{ij} \overline{n_i n_j A} = 0}$ when $A$ does not depends on the filling numbers of sites $i$ and $j$. It leads to the expression
\begin{multline}\label{nl-ij}
\left( \frac{d}{dt} \right)_{ij} \overline{c}_{i\cup j\cup I} = (\overline{n}_i - \overline{n}_j) J_{ij}^{\{ I \}} + \\
J_{ij}\left[ (\overline{n}_i - \overline{n}_j) \overline{c}_I + \overline{c}_{I \cup i} - \overline{c}_{I \cup j} \right].
\end{multline}
Here $J_{ij}$ is the electron flow between sites $i$ and $j$. In terms of the averaged filling numbers and covariations its expression is
\begin{multline}\label{nl-J}
J_{ij} = W_{ij}(1-\overline{n}_i) \overline{n}_j  - W_{ji}(1-\overline{n}_j) \overline{n}_i + \\
 (W_{ji} - W_{ij})\overline{c}_{\{ i,j\}}.
\end{multline}
The currents $J_{ij}$ control the rate equation for the averaged filling numbers as usual
\begin{equation}\label{nl-n}
\frac{d}{dt} \overline{n}_i = \sum_{j \ne i} J_{ij}.
\end{equation}

Equations (\ref{nl-divis}), (\ref{nl-JI-def}), (\ref{nl-ik}), (\ref{nl-ij}), (\ref{nl-J}) and (\ref{nl-n}) compose the non-linear system of correlation kinetic equations.
When all the possible correlations are taken into account the system is exact, however, in this case it includes an extremely large number of variables. When all the correlations are neglected and only averaged filling numbers are considered, it is reduced to the well-known system of mean-field equations. In general case the system can be cut at some point to achieve the balance between correctness and computation possibility.

\section{Significance of the non-equilibrium correlations}
\label{sect-cur}

In the present section we use the correlation kinetic equations to calculate the dependence of correlations on distance and order and the effect of correlations on the currents. The equations are solved numerically for different disordered systems.

We start from a simple system on a square lattice with $10 \times 10$ sites. Only hopping between neighbor sites is allowed. The tunneling integrals between neighbor sites are selected with the distribution $t_{ij} \propto 10^{-x}$ where $x$ is randomly selected in the $(0,1)$ interval. The energies $\varepsilon_i/T$ are randomly selected in the interval $(-3,3)$. It allows us to consider both energy and position disorder.

The results of the calculation in such a system are given in the figure~\ref{fig:SmSys}. In Fig.~\ref{fig:SmSys}(A) the structure of the system, i.e. the distribution of the overlap integrals $t_{ij}$ and energies $\varepsilon_i$ is shown. The overlap integrals are shown with the width of grey lines between neighbor sites and the energies by the colors of the sites. The meanings of the site colors are shown by the color bar. The dependence of the obtained total current on the correlations included into calculations is shown in Fig.~\ref{fig:SmSys}(B). The labels $(r,o)$ on $x$-axis stand for these approximations. $o$ denotes the maximum order of the correlations included into the computation. $r$ is the threshold distance for the correlations along the edges of the lattice. The correlations with a distance larger than $r$ are neglected. For example $x$-axis label $(5,3)$ denotes that the correlation of $3$-d and lower orders were included provided that maximum distance between sites along the lattice edges was not longer than $5$. $(0,1)$ point is the Miller-Abrahams approximation.  $(1,2)$ point is the approximation adopted in \cite{Aleiner1}. The results in Fig.~\ref{fig:SmSys}(B) are averaged over 100 systems with identical statistics of energies and overlap integrals but different disorder realizations. The red line on the figure corresponds to the current calculated with Monte-Carlo algorithm (averaged over the same 100 systems). The figure shows that correlation kinetic equations correctly reproduces the Monte-Carlo simulations provided that enough correlations are included into equations.

The distribution of currents across the system is shown in Fig.~\ref{fig:SmSys}(C). Fig.~\ref{fig:SmSys}(D) shows the difference in this distribution between Miller-Abrahams calculation and the calculation including long-range correlations corresponding to the point $(5,4)$ in Fig.~\ref{fig:SmSys}(B). The results provided in Fig.~\ref{fig:SmSys}(B) and (D) indicate that the effect of correlations both on the conductivity and current distribution is relatively small ($\sim 10\%$) for the considered parameters.

\begin{figure}[htbp]
   \centering
       \includegraphics[width=0.4\textwidth]{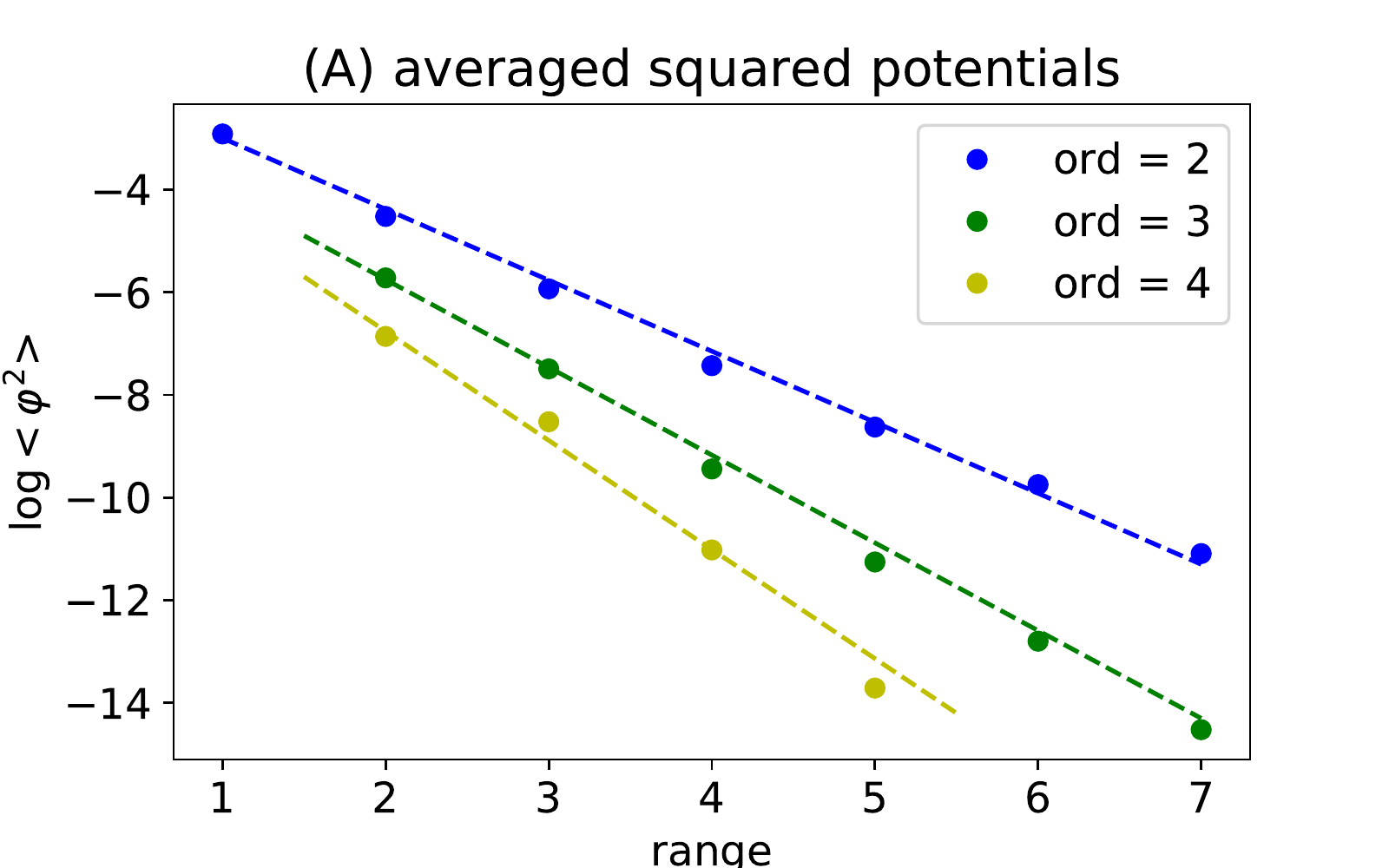}
       \includegraphics[width=0.4\textwidth]{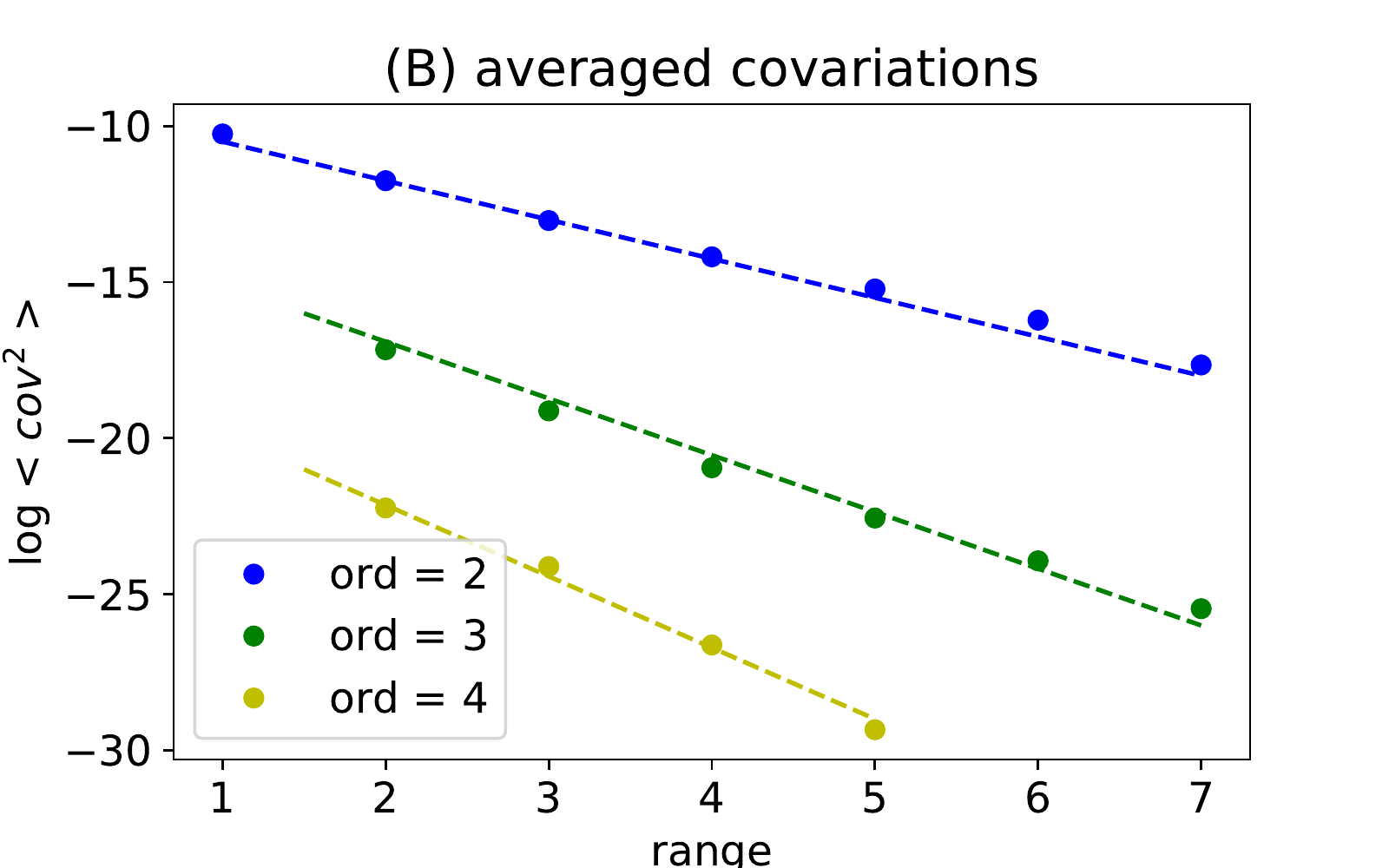}
       \caption{The dependence of correlations on distance and order. (A) averaged squared potential $\varphi_I$, (B) averaged covariations.}
   \label{fig:avCov}
\end{figure}

\begin{figure*}[htbp]
   \centering
       \includegraphics[width=1.0\textwidth]{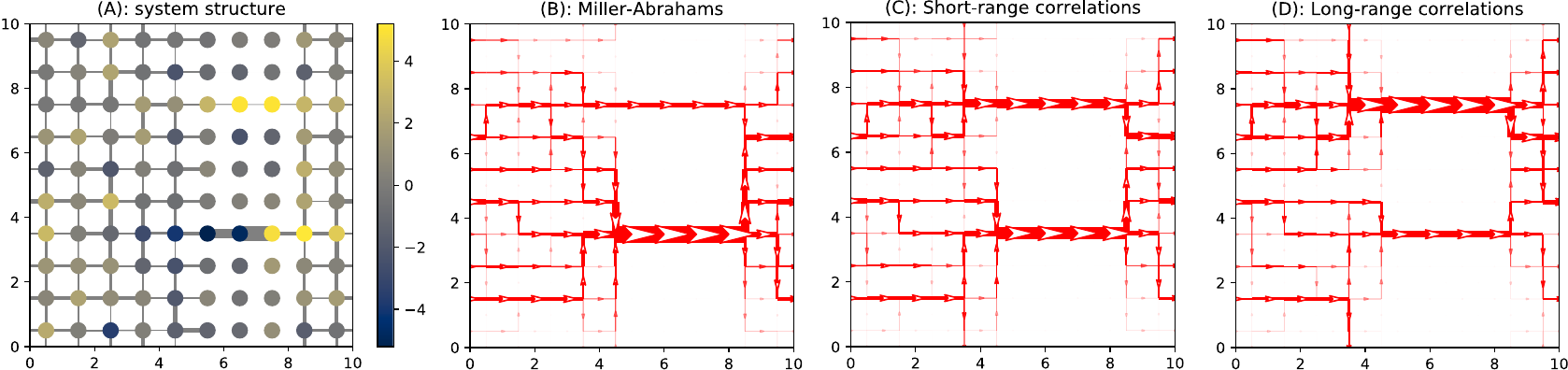}
       \caption{A special system with a strong effect of correlations on currents. (A) the structure of the system. (B)-(D) the current distributions calculated in different approximations as described in the text. }
   \label{fig:SpSys}
\end{figure*}

Figure \ref{fig:avCov} shows how the correlations decrease with order and intersite distance. In Fig.~\ref{fig:avCov}(A) we show correlation potentials $\varphi_I$ and in Fig.~\ref{fig:avCov}(B) we show the covariations. Different colors correspond to the different orders of correlations. The $x$-axis corresponds to the maximum distance between sites in the considered correlation along the lattice edges. Points are numerical results averaged over 100 random system. Dashed lines correspond to the exponential decay.

Both potentials and covariations decay with distance and order. It justifies the cutoffs in the correlation kinetic equations. The decay of covariation with order is much faster than the decay of potentials due to the product $\prod_i n_i^{(0)}(1-n_i^{(0)})$ in Eq. (\ref{udef}).

How do the results provided in Fig.~\ref{fig:SmSys} indicating the relatively small significance of correlation agree with the result of \cite{Aleiner1} stating that a resistor with different signs of site energies can grow exponentially due to the correlations? We believe that what is important is the rate of correlation relaxation due to its transfer to other sites of the system. In \cite{Aleiner1} this rate was assumed to be comparable to the rate of electron transition in the critical resistor. However, usually the critical resistor that determines the conductivity of the disordered system is connected to the rest of the system with much more conductive resistors. It means that the correlation can leave the critical resistor quite fast preventing a large increase of the resistivity. To demonstrate this idea we construct a system where the correlations leave the critical resistor slowly and their effect should be strong.

The system is shown in Fig.~\ref{fig:SpSys}(A). The notations on this figure are the same as in Fig.~\ref{fig:SmSys}(A). The system is composed of two reservoirs with relatively good conductivity connected by the two bridges with small conductivity. However, the reasons for the small conductivity of bridges are different. In the upper bridge the overlap integrals $t_{ij}$ are relatively small but the energies of the sites have the same sign. In the lower bridge $t_{ij}$ are larger and the small conductivity is controlled by the hop between two sites with large energies with opposite signs. However, the structure of the bridge, i.e. the relatively small overlap integrals on the sides of the bridge, prevents the correlations from leaving the bridge too fast.

In Fig.~\ref{fig:SpSys}(B) we show the distribution of currents in this system calculated in Miller-Abrahams approximation. In this approximation the conductivity of the lower bridge is higher than the conductivity of the upper one. Most of the current is concentrated on this bridge. Figure \ref{fig:SpSys}(C) corresponds to the approximation used in \cite{Aleiner1}, when only pair correlations on the neighboring sites are taken into account. In this case the calculated currents split between the bridges.  Figure \ref{fig:SpSys}(D) shows the current distribution calculated with taking into account the correlations up to the fourth order and with the distance between sites along the bonds of the lattice no longer than $5$. In this case the resistivity of the lower bridge is $\sim 10$ times larger than in Miller-Abrahams approximation. Most of the current is concentrated on the upper bridge in Fig.~\ref{fig:SpSys}(D).

\begin{figure}[htbp]
   \centering
       \includegraphics[width=0.4\textwidth]{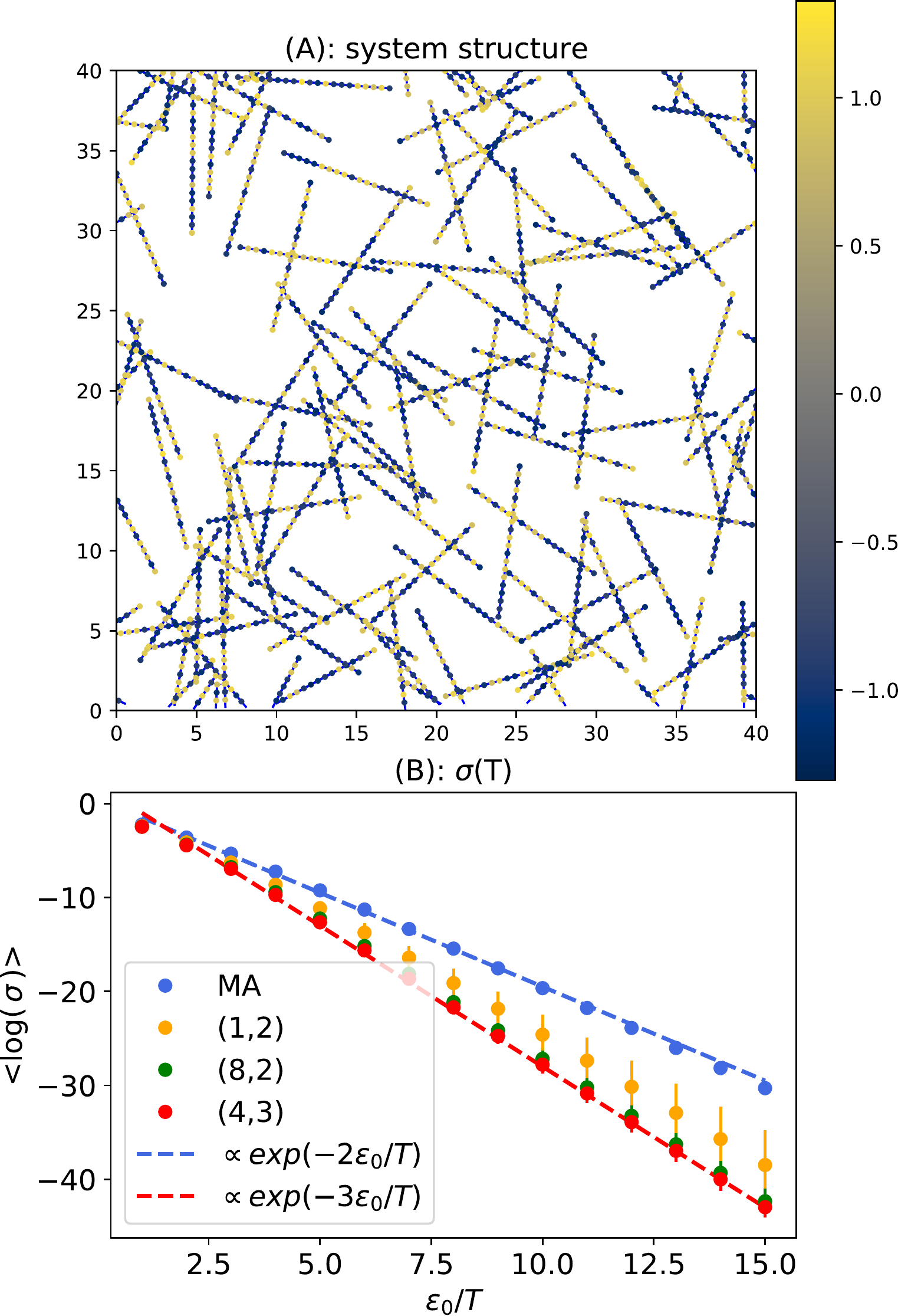}
       \caption{(A) the sample composed by the polymer chains. (B) the results for this sample.}
   \label{fig:Poly}
\end{figure}

The results provided in Fig.~\ref{fig:SpSys}  prove that the correlations can be important for certain arrangements of site energies and overlap integrals. The arrangement in Fig.~\ref{fig:SpSys} cannot be considered as disordered. However, this system helps to understand the conditions that lead to strong contribution of the correlations to conductivity. First of all the conductivity of the system should be controlled by the hops between sites with different sign of $\varepsilon_i - \mu$. It ensures that current effectively generates the correlations in the pairs responsible for resistivity. These correlations should not be able to relax easily due to the transport to different parts of the system. Ideally in a pair $i-j$ important for the resistivity of the whole system both the sites $i$ and $j$ should be parts of small groups that cannot be leaved by the charge carriers easily. The possible ``borders'' preventing fast transition of the carrier outside such groups are the hops between sites with different sign of energy or hops with small overlap integrals $t \ll t_{ij}$.

In Fig. \ref{fig:Poly} we provide results for a random system where these conditions are met. The system is composed by short chains of 20 sites that randomly have positive energy $+\varepsilon_0 + \delta\varepsilon$ or negatibe energy $-\varepsilon_0 + \delta\varepsilon$. Zero energy corresponds to the chemical potential. $\delta\varepsilon$ is a random energy with gaussian distribution with the width equal to $0.1\varepsilon_0$.  The overlap integrals inside a chain are relatively large. Where the chains overlap the inter-chain hopping is possible. Its rate is 10 times smaller than the rate of hopping inside a chain. Positions of chains in the numeric sample are random. The site energies and overlap integrals of such a system are shown on Fig.~\ref{fig:Poly}(A).  We believe that such a system can be realized in polymers randomly composed from two kinds of monomers.

In Fig. \ref{fig:Poly}(B) we show the temperature dependence of conductivity calculated in such systems in the four approximations: the Miller-Abrahams (MA) approximation, the approximation $(1,2)$ where only pair correlations in close pairs are taken into account and two approximations $(8,2)$ and $(4,3)$ that take into account the correlations on distant sites. The logarithm of conductivity shown on \ref{fig:Poly}(B) is averaged over 100 disordered systems with similar statistics of site energies and overlap integrals. The points are numerical results. The vertical bars show the standard deviation of $\ln(\sigma)$ in the ensemble averaging. The dashed lines correspond to the Arrhenius law.

In all the approximations the temperature dependence of the conductivity agrees with the Arrhenius law $ {\sigma \propto \exp(- \varepsilon_A/T)}$. However the activation energies $\varepsilon_A$ calculated in different approximations are different.
$\varepsilon_A$ calculated in Miller-Abrahams approximation is equal to $2\varepsilon_0$. It is the energy of the phonon required for the hop from site with energy $-\varepsilon_0$ to the site with $\varepsilon = +\varepsilon_0$. Both the approximations $(8,2)$ and $(4,3)$ lead to the activation energy $3\varepsilon_0$. It is the result of non-equilibrium correlations. Similar result was obtained in \cite{Richards} for infinite one-dimensional (1D) chain where sites with positive and negative are arranged in alternating order. Here we show that this result holds for random composition of chains and for quasi-1D samples composed of finite chains.

The activation energy obtained in the approximation $(1,2)$ that takes into account only the close-range correlations is between $2\varepsilon_0$ and $3\varepsilon_0$. Note that the results in $MA$, $(8,2)$ and $(4,3)$ approximations are reproduced quite well in different disordered samples (the standard deviations of $\ln(\sigma)$ shown with vertical bars in Fig \ref{fig:Poly}(B) are small). Nevertheless it is not the case for $(1,2)$ approximation. In some samples $(1,2)$ approximation reproduces mean-field results while in others it reproduces the results of approximations that take long-range correlations into account. While our numerical samples are large compared to the single chain, they are still mesoscopic. The conductivity of a sample is controlled by a small number of important chains that ensure the percolation. The distribution of sites in these chains determine if $(0,2)$ approximation gives the correct result for conductivity or not.

\subsection{Correlations in the Mott law regime}
\label{sect-cur-mott}

Two different opinions exist on the importance of the non-equilibrium correlations for the hopping transport in the materials in Mott law regime. Levin, Nguen and Shklovskii \cite{Shklovskii-cor} calculated the temperature dependence of conductivity in a cube with 1000 sites in this regime with Monte-Carlo simulation and within Miller-Abrahams resistor model. The first method takes the correlations into account while the second method does not. The difference between the currents obtained by the methods was of the order of unity and was considered negligible compared to the exponential dependence of the current on temperature. Agam and Aleiner \cite{Aleiner1} pointed out that the resistance between two sites with different signs of energy can grow exponentially due to the correlations. It was predicted that at low temperature two parallel percolation networks should exist. If it is the case the distribution of currents in the system should be totally different from the one predicted by the Miller-Abrahams network. The failure of the Mott law was also predicted in \cite{Aleiner1} for the sufficiently low temperatures.

\begin{figure*}[htbp]
   \centering
       \includegraphics[width=1.0\textwidth]{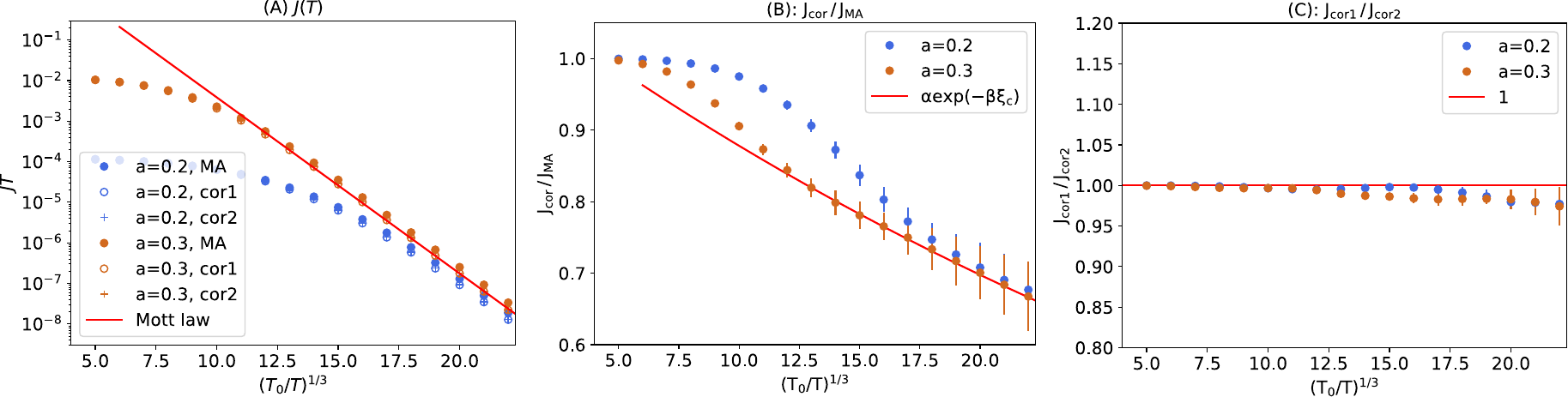}
       \caption{Current in the Mott law regime calculated in three approximations: the Miller-Abrahams approximation (``MA''),
       approximation where pair correlations up to the distance $2 r_c$ are taken into account (``cor1'') and with correlations up to the third order and up to the distance $r_c$ (``cor2''). (A) the currents for two localization lengths compared to the Mott law, (B) the difference between ``MA'' approach and ``cor1'' approach, (C) the difference between ``cor1'' and ``cor2'' approaches. In figures (B) and (C) points correspond to disorder-averaged values and bars to the standard deviation of this averaging. }
   \label{fig:mott}
\end{figure*}

 To model a material in the Mott law regime we consider systems with 5000 sites randomly placed in square numerical samples. The site concentration is equal to unity and determines the units of distance. The overlap integrals between the sites $i$ and $j$ depend on the distance between sites $r_{ij}$,
\begin{equation}\label{tij}
t_{ij} = t_0 \exp(-r_{ij}/a).
\end{equation}
 Here $a$ is the localization length. We consider two values $a=0.2$ and $a=0.3$ to model different degrees of positional disorder. The energies of sites were randomly selected in the interval $(-1,1)$ that determine the units of energy. Therefore the energy disorder was controlled by temperature $T$. $T \ll 1$ stands for the strong energy disorder. The Mott law should be observed at sufficiently low temperatures.

The crucial parameter in this regime is the value of the critical exponent $\xi_c = (T_0/T)^{1/3}$. Here $T_0 = \beta_{2D}/ga^2$. $\beta_{2D} \approx 13.8$ is the numeric coefficient and $g$ is the density of states at the Fermi level \cite{Efr-Sh}. This critical exponent determines the maximum distance of the hop contributing to the current $r_c = a\xi_c/2$ and the interval of energies $-\varepsilon_{\rm max} < \varepsilon_i < \varepsilon_{\rm max}$, $\varepsilon_{\rm max} = \xi_c T$ that includes all the sites relevant for conductivity. To simplify the numeric calculations we excluded from calculations all hops with distance longer than $a(\xi_c + 3)/2$ and all the sites with $|\varepsilon_i| > (\xi_c+3) T$. We solved the system of equations (\ref{ddt}), (\ref{dtl-ik}), (\ref{dtl-ij}) in three approximations: in the Miller-Abrahams approximation (MA), in the approximation when pair correlations up to the distance $2r_c$ are taken into account (we denote it as ``cor1'' approximation) and in the approximation when correlation up to the third order and distance $r_c$  are calculated (``cor2'' approximation).
The results are averaged over ten disorder configurations.

The results of our simulation are shown in Fig.~\ref{fig:mott}. Figure \ref{fig:mott}(A) is the comparison between the Mott law and the numeric results. In the Mott law, the exponential part of the temperature dependence of conductivity is $\exp(-\xi_c)$. The pre-exponential part is governed by system dimensionality and the pre-exponential part of dependence of hopping rates on energies and distances \cite{Efr-Sh}. In our case it is determined by Eqs.~(\ref{Wij}) and (\ref{tij}) and leads to pre-exponential term $1/T$ in the temperature dependence of the total current $J(T)$. Therefore $T\times J$ should follow the universal law $TJ = const\times \exp(-(T_0/T)^{1/3})$ independent on localization distance $a$ when the system is in the Mott law regime. Figure \ref{fig:mott}(A) shows that the Mott law regime is achieved for $a=0.3$ when $\xi \ge 12$ and for $a=0.2$ when $\xi \ge 15$.

Figure \ref{fig:mott}(B) shows the ratio of the current $J_{cor}$ calculated in cor1 approximation to $J_{\rm MA}$ calculated in Miller-Abrahams approximation. The points on the figure shows the value $J_{cor}/J_{\rm MA}$ averaged over 10 numerical samples. The vertical bars show the standard deviation of $J_{cor}/J_{\rm MA}$. This deviation is due to the finite size of numerical samples.  The ratio $J_{cor}/J_{\rm MA}$ decreases with increasing $\xi_c$, however, it does so rather slowly and stays $\sim 1$ in all the range of the parameters considered. This ratio also should follow some universal dependence on $\xi_c$ in the Mott law regime. The figure shows that in this regime the results for $a=0.2$ and $a=0.3$ are similar. We compare the decrease with exponential law $\alpha\exp(-\beta\xi_c)$ with $\alpha \approx 1.1$ and $\beta \approx 0.02$. Although the agreement is achieved the small magnitude of decrease of $J_{\rm cor}/J_{\rm MA}$ with $\xi_c$ does not allow to reliably extract the asymptotic at large $\xi_c$ from our calculations for $\xi_c \le 22$.

In Fig.~\ref{fig:mott}(C) we show the ratio of the currents calculated with different correlations included, in the approximations cor1 and cor2. This ratio is close to unity. We believe that it indicates that the both approximations correctly describe the effect of correlations on the transport in the considered systems.

\begin{figure}
   \centering
       \includegraphics[height=0.91\textheight]{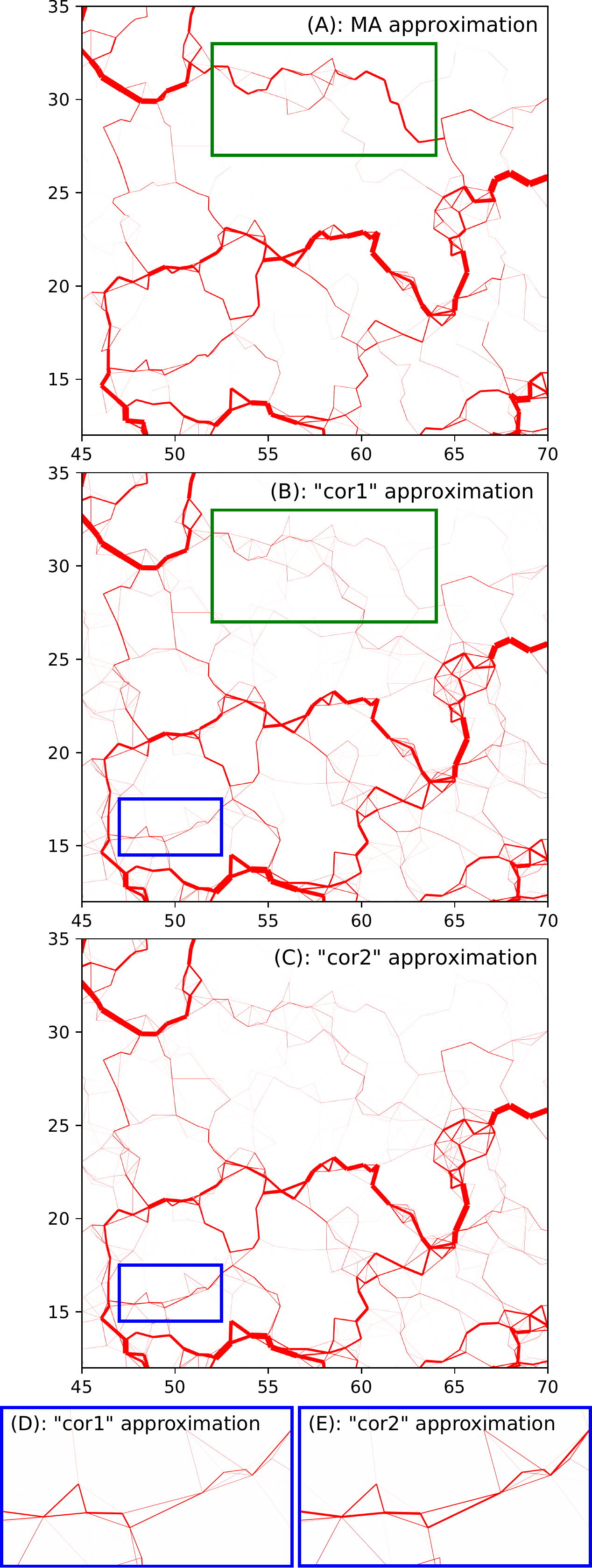}
       \caption{Distribution of current in a system in Mott law regime calculated in different approximations: Miller-Abrahams approximation (A), ``cor1'' approximation (B) and ``cor2'' approximation (C). Panels (D) and (E) show zoom of blue (bottom-left) rectangles in  panels (B) and (C) correspondingly.}
   \label{fig:distrib}
\end{figure}

In Fig.~\ref{fig:distrib} we show the distributions of currents calculated in the three approximations in some part of one of our numeric samples. We considered $a=0.2$ and $(T_0/T)^{1/3} = 20$, i.e. the sample is nearly as deep in the Mott law regime as it is possible with our computation capabilities. The difference between the distribution of currents in Miller-Abrahams approximation and in ``cor1'' approximation is clearly visible. The part of the sample where this difference is the strongest is marked by a green (upper) rectangle. However, the current distributions still have similar patterns. We do not observe the two percolation networks predicted in \cite{Aleiner1}.
The difference in current distribution between ``cor1'' and ``cor2'' approximation is much less pronounced. However, some differences can be noted in the area of the sample marked with a blue (bottom-left) rectangle. Area in these rectangles is shown in larger scale in Fig.~\ref{fig:distrib}(D,E).  In supplementary materials \cite{sup} we provide similar current distribution pictures for the whole area of 10 numeric samples. All the samples have the parameters $a=0.2$ and $\xi_c = 20$.

The Mott law is observed in a number of quite different materials. When it is observed the transport is controlled by a small number of sites with energies close to the Fermi level. In this case the overlap integrals usually can be estimated with eq. (\ref{tij}) with some localization length $a$. Also the density of states at the temperatures corresponding to Mott law can be considered as a constant. All the results that follow from this statistics should be universal for all the materials that exhibit Mott law. Therefore, the relative contribution of non-equilibrium correlations to the conductivity in all such materials should be the same.

\subsection{Correlations far from equilibrium}
\label{sect-cur-nl}

In this section we calculate the non-equilibrium correlations of filling numbers in strong applied electric field when the linear-response approximation is not valid. We discuss the effect of correlations on transport in systems with two types of disorder: position disorder related to random overlap integrals $t_{ij}$ and energy disorder related to random energies $\varepsilon_i$ of sites. In most of materials both types of disorder coexist. However, in principle they can exist and be modeled separately. When the temperature is larger than the width of energy distribution  the energy disorder is not important for transport. However, if the site concentration and localization length are small $n^{1/2}a \ll 1$ the system is still strongly disordered. This regime is known as nearest neighbor hopping \cite{Efr-Sh}. One can also consider the opposite case when the sites are positioned on some lattice and all the overlap integrals are the same. The system is disordered only due to the distribution of site energies that should be wider than temperature.

\begin{figure*}[htbp]
   \centering
       \includegraphics[width=0.8\textwidth]{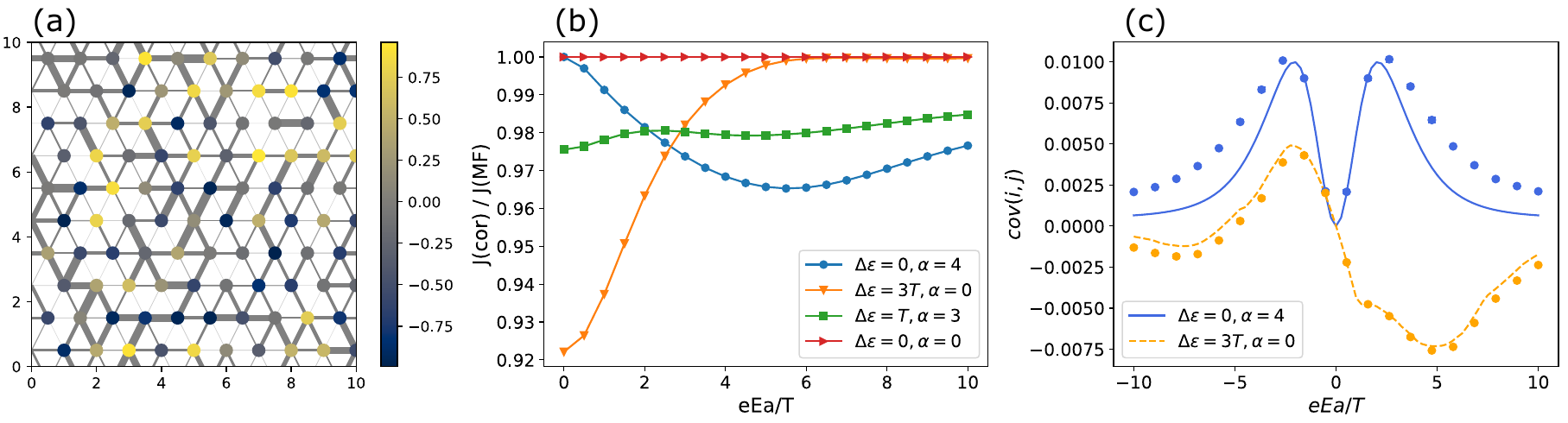}
       \caption{The correlations of filling numbers in high electric field. (a) the structure of considered system. (b) The ratio of current calculated with correlation to the current calculated in the mean-field approximation as a function of electric field. The results are averaged over 100 disordered samples (c) The dependence of covariation of filling numbers in close pairs on the applied field. }
   \label{fig:nonlin}
\end{figure*}

If only the position disorder is present in the system, the correlations of filling numbers do not appear and do not affect the transport properties at low electric field. It can be understood from Eq. (\ref{J2}). The current $\widetilde{J}_{ij}$ responsible for generation of correlations is proportional to the difference of equilibrium filling numbers $n_i^{(0)} - n_j^{(0)}$. This difference is equal to zero without energy disorder. It leads to the absence of correlations of filling numbers and their effect on the current in the linear response regime. When the applied field is strong the position disorder leads to different occupation numbers of different sites and the correlations can appear and affect the currents. The correlations in this case are proportional to $E^2$ at small applied fields $E$.

The situation in the system with energy disorder is reversed. At small applied fields the correlations are proportional to $E$ and affect the current in linear response regime. However, in strong field $E$ their effect on the current is suppressed. It can be understood with the following arguments. When the voltage on the neighbor sites is strong compared to the temperature and energy disorder the hops can be divided into two types. The hops in the direction of electric field occur with phonon emission and in our approximations (see Eq.~(\ref{Wij})) are not sensitive to the energies of involved sites. The backward hops against the field are impossible and again are not sensitive to the site energies. Therefore, the energy disorder is  not important in the limit of a strong applied electric field. If no position disorder is present in the system it behaves like an ordered one and displays no effect of correlations on transport. However, the above arguments do not take into account the hops exactly perpendicular to the electric field. When the  transport involves a significant number of these hops (it occurs, for example, in the system shown in Fig.~\ref{fig:SmSys}) the correlations can affect transport in system with energy disorder in a high electric field.

To simulate the difference between the two types of disorder we model a system of sites on the triangle lattice. The electric field is applied along x-axis. The energies of the sites are uniformly distributed in the interval $(-\Delta \varepsilon, \Delta \varepsilon)$. The overlap integrals between neighbors are equal to $t_{ij} = t_0 \exp(-\alpha x_{ij})$ where the values $x_{ij}$ are uniformly distributed in $[0,1]$ interval. The parameter $\alpha$ controls the disorder in overlap integrals that represents the position disorder in our model. $\Delta \varepsilon$ controls the degree of energy disorder. The temperature is considered to be equal to unity $T=1$.  The triangle lattice was selected to exclude hops that are perpendicular to the applied electric field.

The considered system with $\alpha=3$, $\Delta \varepsilon = 1$ is depicted in Fig.~\ref{fig:nonlin}(a). In Fig.~\ref{fig:nonlin}(b) we show the difference between current $J_{\rm MF}$ calculated in the mean-field approximation and current $J_{\rm cor}$ calculated with correlations of the second order taken into account for sites with distance along lattice edges up to 5. We provide the results for the four types of disorder. For the energy disorder (yellow curve, $\alpha = 0$, $\Delta \varepsilon = 3T$) $J_{\rm cor}<J_{\rm MF}$ at low field $eEa \ll T$. Here $a$ is the length of the lattice edge. However, at large fields $eEa \gg T$, the currents are equal, $J_{\rm cor}/J_{\rm MF} = 1$. It means that correlations of filling numbers do not affect the transport. The situation is different for position disorder $\Delta \varepsilon = 0$, $\alpha = 4$, blue curve. $J_{\rm cor}/J_{\rm MF} = 1$ at small field $E$ indicating that correlations are not important in the linear response regime. However, at strong field the effect of correlations appears. When the disorder has a mixed type ($\Delta \varepsilon = 1$, $\alpha = 3$, green curve) the effect of correlations exist for all the electric fields. Finally, the red curve corresponds to an ordered system with $\alpha=0$, $\Delta \varepsilon = 0$. The correlations are absent and have no effect on the transport in an arbitrary electric field in this case.

In Fig.~\ref{fig:nonlin}(c) we show typical dependence of covariations of filling numbers on neighboring sites on the electric field. Lines in this figure correspond to CKE calculation. In this calculation the pair correlations were taken into account for distance between sites $\le 6$ and the correlations of the third order for distance between sites $\le 3$. The distance was measured along the lattice bonds. The points correspond to Monte-Carlo simulation.

In a system with energy disorder, the dependence of covariation on electric field is linear at low field and asymmetric with respect to reversion of the field direction. In a system with position disorder the covariations in a small field are proportional to $\propto E^2$ and saturate at strong fields.

\begin{figure*}
   \centering
       \includegraphics[width=1.0\textwidth]{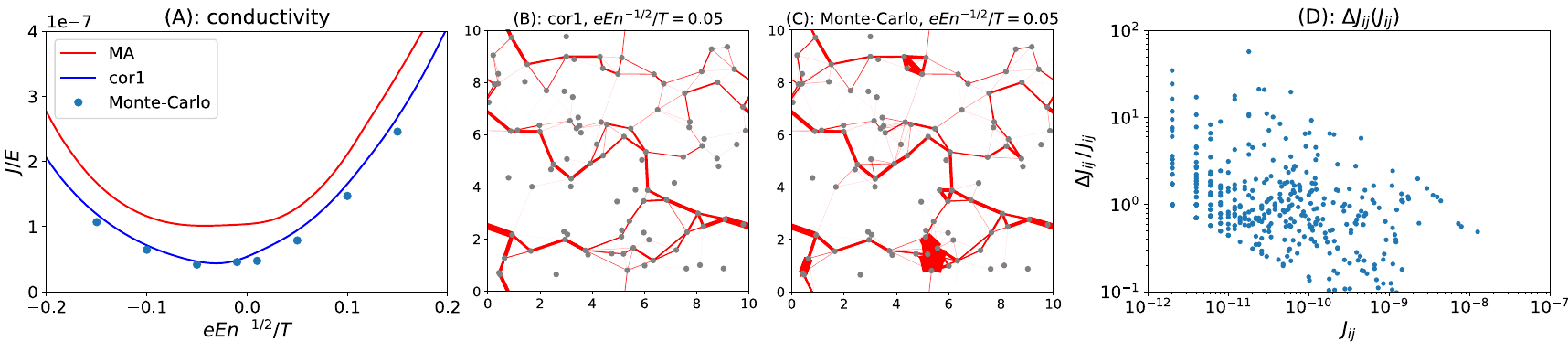}
       \caption{The comparison between CKE calculations and Monte-Carlo algorithm. (A) the dependence of conductivity on the applied field calculated in Miller-Abrahams (MA) approximation, in ``cor1'' approximation and with Monte-Carlo algorithm. (B) The current distribution calculated in ``cor1'' approximation. (C) The current distribution calculated with Monte-Carlo algorithm. (D) the averaged current between sites and their standard deviations in Monte-Carlo algorithm.}
   \label{fig:MC}
\end{figure*}

The Monte-Carlo simulation and CKE calculation agree at relatively small electric field in Fig.~\ref{fig:nonlin}(c). However at strong field there is a disagreement in estimate of correlations. It is especially pronounced for position disorder. Nevertheless the currents calculated with the two methods agree with precision $\sim 1 \%$. It seems that in a system with position disorder in strong electric field large number of correlations should be included into the calculation to reliably estimate the correlations in close pairs.

\section{Comparison with the Monte-Carlo simulation}
\label{sect-monte}

In this section we compare the proposed method of correlation kinetic equations with the Monte-Carlo simulations that are conventionally used when the mean field approximation is not sufficient. In Fig. \ref{fig:SmSys} we have shown that Monte-Carlo simulation agrees with CKE provided that enough correlation are included into the equations. Therefore the question of applicability of CKE is the question ``is the number of included correlations sufficient?''.

In this section we compare the methods for materials in Mott law regime. We were not able to reproduce all the results of Sec. \ref{sect-cur-mott} with Monte-Carlo simulation. Accordingly we focus on the small system with the site statistics described in  Sec. \ref{sect-cur-mott} and $a=0.2 n^{-1/2}$, $\xi=20$. We compare ``cor1'' approximation with Monte-Carlo simulations.

The results of this comparison are shown in Fig.~\ref{fig:MC}. On Fig.~\ref{fig:MC}(A) we show the dependence of non-linear conductivity $J/E$ on the applied electric field calculated with the three methods. Red curve is calculated in Miller-Abrahams approximations, blue curve is calculated in ``cor1'' approximation. The blue points correspond to the Monte-Carlo simulation. The time of Monte-Carlo simulation was $\sim 20$ times larger than the time of CKE calculations. The difference between results of Miller-Abrahams network and the more precise methods is significant $\sim 2$ times. The difference between ``cor1'' approximation and Monte-Carlo simulation is much smaller but is observable. It is related to the correlations neglected in ``cor1'' approximation.

In Fig.~\ref{fig:MC}(B) and Fig.~\ref{fig:MC}(C) we show the current distribution calculated for $eEn^{-1/2}/T = 0.05$ with ``cor1'' approximation and Monte-Carlo simulation correspondingly. The current distribution calculated with ``cor1'' approximation looks reasonably. The distribution on Fig.~\ref{fig:MC}(C) has several unexpectedly large currents that break the charge-conservation law or form circular currents. These artificial currents appear because of fluctuations in Monte-Carlo algorithm. Note that statistical error of conductivity calculated with Mote-Carlo algorithm at these field is $\sim 2\%$. The time of Monte-Carlo calculation was sufficient to reliably calculate the conductivity but the current distribution obtained by this algorithm still is unreasonable.
At the smaller field $eEn^{-1/2}/T = 0.01$ that is close to the linear response regime the current distribution obtained with Monte-Carlo simulation is even much less accurate.

In Fig.~\ref{fig:MC}(D) we show the standard deviations of local currents $\Delta J_{ij}$ calculated with Monte-Carlo algorithm versus their averaged values. The standard deviations of most of the small currents are usually larger or comparable with their averaged values. It indicates that these currents are not reliably estimated with our simulation. Even among the large currents there are ones with large standard deviations $\Delta J_{ij} > J_{ij}$.

The small electric field (i.e. the linear-response regime) requires the largest times to calculate with Monte-Carlo algorithm because the current fluctuations are large compared to field-induced currents. On the other hand it is the best situation for CKE approach. In this case the correlation equations are linear and can be solved more effectively than the non-linear equations that should be applied far from equilibrium. To our opinion the CKE approach is usually more efficient than Monte-Carlo at small electric field. At strong electric fields Monte-Carlo sometimes appears to be more useful than CKE especially when the detailed information on the system (the current distribution) is not required.

\section{Discussion}
\label{sect-dis}

In the linear response regime, the non-equilibrium correlations always decrease the current in comparison to one calculated with Miller-Abrahams approximation. It can be proved rigorously. The linear correlation kinetic equations are equivalent to the following minimization problem
\begin{multline}
F= \frac{1}{2}\sum_{\substack{I,i,k\\i<k}}\Gamma_{ik}^{\{I\}} \times \\
\left[\overline{u}_{k\cup I} - \overline{u}_{i \cup I} + s_{ij}^{\{I\}} + \left(n_k^{(0)} - n_i^{(0)} \right) \overline{u}_{i \cup k \cup I}\right]^2.
\end{multline}
Here $F$ is the function of potentials $\overline{u}_I$ that should be minimized to find the correct potentials. $I$ is any set of sites including $\emptyset$ that does not contain sites $i$ and $k$.
The conditions of minimum $\partial F/\partial \overline{u}_{I}=0$ yield the system of Kirchhoff equations (\ref{eq:Kirchhoff}).

The term $(n_k^{(0)} - n_i^{(0)}) \overline{u}_{i \cup k \cup I}$ in the definition of $F$ is responsible for the interconnection of levels of the equivalent circuit (Fig.~\ref{fig1} (b)). It simultaneously leads the to upward interconnection, i.e the currents $\widetilde{J}_I$ that flow from ground to the upper levels, and to downward interconnection, i.e. the effect of high-order correlations on the currents of lower-order ones.  Without this term, the minimization of $F$ yields a system of Kirchhoff equations for the Miller-Abrahams resistance network.

The current flow in the system is proportional to the function $F$ calculated with correct potentials $\overline{u}_I$ that satisfy the Kirchhoff equations. It can be shown that the derivative $\partial F/\partial E = eJL/T$. Here the derivative $\partial F/\partial E$ is taken with constant potentials $\overline{u}_I$. $J$ is the total particle flow in the system. $L$ is the system size. $J$ is proportional to electric field and $F$ is a bi-linear function of the field. It leads to the expression
\begin{equation}
    J = \frac{2T}{eEL}\min_{\{\overline{u}\}}F.
\end{equation}
Here $\min_{\{\overline{u}\}}F$ is the minimum of the function $F$ at given field $E$.

When we cut the system of kinetic equations we artificially set some potentials to be equal to zero. It imposes additional conditions to minimum $\min_{\{\overline{u}\}}F$ and can only increase it. Therefore, when one neglects some correlations the current becomes overestimated. And vice versa: each additionally included correlation decreases the calculated current.

This result, however, is valid only in linear response regime. When the electric field is strong the result depends on system configuration. In some configurations the mean-field approximation overestimates the current while in others it underestimates the current.

\begin{figure}[htbp]
   \centering
       \includegraphics[width=0.45\textwidth]{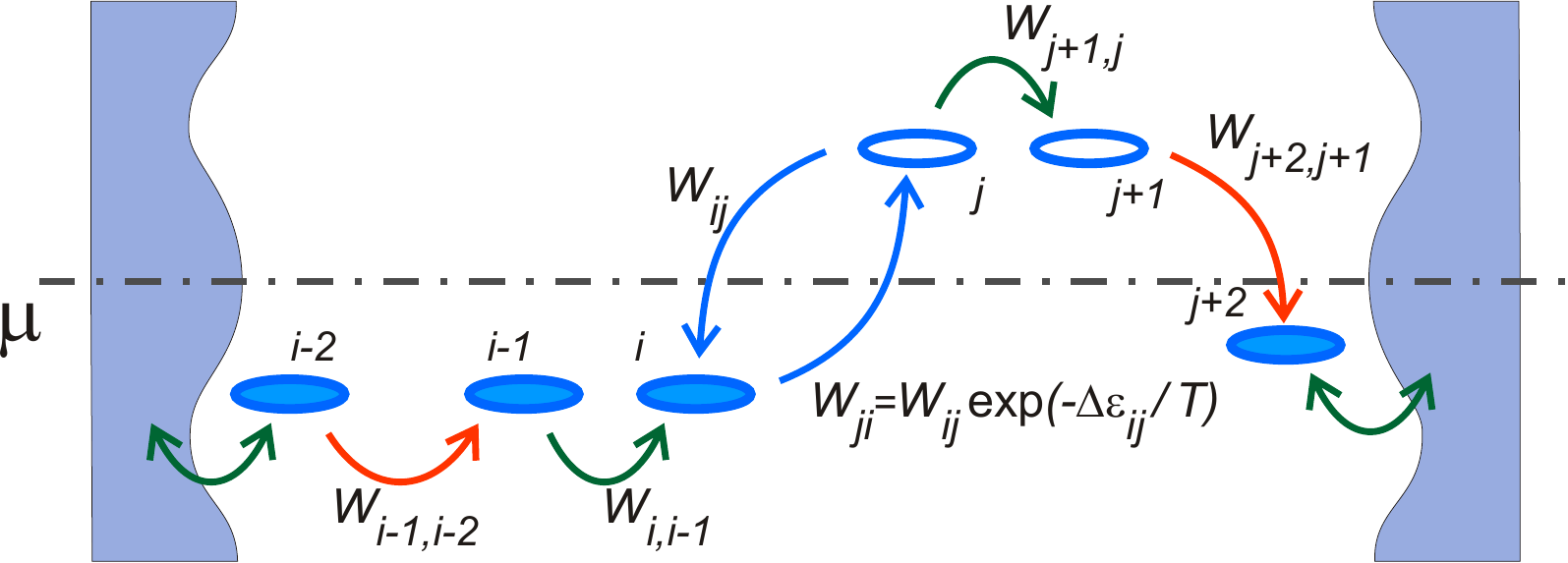}
       \caption{The chain with resistance dominated by the hop between sites with different signs of energy.}
   \label{fig:dis}
\end{figure}

The reason for the decrease of current due to correlations is the fast recombination of electron-hole pairs that appear after the hop from a site with negative energy to a site with positive energy. Consider a chain of sites
where the conductivity is controlled by the hops $i \leftrightarrow j$ between sites with different signs of energy (Fig.~\ref{fig:dis}). The rate of $i \rightarrow j$ hops is controlled by the small value $W_{ji}$. The backward hops $j \rightarrow i$ have a much higher rate $W_{ij} \gg W_{ji}$ but occur only when site $i$ is free and site $j$ is filled. Both of the events are quite improbable. The current $i \rightarrow j$ makes the hops $i \rightarrow j$ more frequent than hops $j \rightarrow i$. After such a hops the electron appears on site $j$ simultaneously with hole on site $i$.
 It leads to additional probability to find free site $i$, filled site $j$ and (what is important) to the positive correlation of the two events. The appeared electron-hole pair can recombine with fast rate $W_{ij}$. It creates ``negative feedback'' for the hopping between sites $i$ and $j$ and can decrease the conductivity of the chain compared to the prediction of the Miller-Abrahams theory.

There is an important condition for this decrease of conductivity to be significant: the electron-hole pair should stay on the pair of sites $i-j$ at least for the time $1/W_{ij}$. It is so when
the hops $j \rightarrow j+1$ and $i-1 \rightarrow i$ are slow compared to the hop $j \rightarrow i$, $W_{i,i-1},W_{j+1,j} \ll W_{ij}$. Otherwise if at least one of this hops is fast the correlation $i-j$ can be efficiently transferred to the other sites. In this case its effect will be negligible at least in the theory that includes only the correlations in close pairs of sites. This condition was understood in \cite{Shklovskii-cor} with analytical solution of the model of four sites.

However if some other hop in the chain is slow the correlation can return to the initial sites. For example if $W_{i-1,i-2} \ll W_{ij}$ the hole will stay on the pair of sites $i-1,i$ for quite a long time.
There is also the second reason for the electron-hole pair to stay close to the initial sites. Let us consider the site $j+2$ to have negative energy. In this case even if the hopping rates $W_{j+1,j}$ and $W_{j+2,j+1}$ are high the electron cannot leave sites $j$ and $j+1$ until a hole appears on site $j+2$.  However, the correlations at some distance should be taken into account to catch such effects.

Finally what is important for the correlation transfer is the branching of the hopping path. Consider that the correlation was created on the pair $i-j$ and that the site $j$ has large number of neighbors. If the rate of transition to at least some of these neighbors is fast there is a high possibility for the electron to leave the site $j$ without recombination with the hole on the site $i$. Therefore the effect of correlations on currents should be the strongest in one-dimensional or quasi-1D systems.

In conclusion we developed the system of kinetic equations that describes non-equilibrium correlations of site filling numbers up to arbitrary order. In the linear response regime it can be reduced to an equivalent circuit that generalizes the Miller-Abrahams resistor network. We show that correlations decrease with order and distance between sites. It allows to cut the system and achieve the balance between correctness and computation capabilities. With our approach we show that in some disordered materials (e.g. in quasi-1D systems with two kinds of sites) the effect of correlations on the current distribution is significant.  However it is not the case for systems with Poisson distribution of sites in Mott law  regime. In such systems the contribution of correlations to currents is of the order of unity for $(T_0/T)^{1/3} \lesssim 20$. The correlations are absent in ordered systems. The different types of disorder have different effects. Energy disorder makes correlations important in small applied electric fields while the position disorder makes them important in strong applied fields.

We are grateful to Y.M.~Galperin, V.I.~Kozub and B.I.~Shklovskii for many fruitful discussions. We acknowledge partial support from RFBR, grant no.~19-02-00184 A.


\begin{thebibliography}{99}


\bibitem{MA}
A. Miller and E. Abrahams, Phys. Rev. 120, 745 (1960).

\bibitem{Efr-Sh}
B. I. Shklovskii and A.L. Efros, "Electronic Properties of Doped
Semiconductors" (Springer, Berlin, 1984).


%%%%------------------ Работы про кулоновские стекла -------------------------------

\bibitem{Col-glass-1}
J.H. Davies, P.A. Lee, T.M. Rice, Phys. Rev. Lett. {\bf 49} 758 (1982)

\bibitem{Col-glass-2}
J.H. Davies, P.A. Lee, T.M. Rice, Phys. Rev. B {\bf 29} 4260 (1984)

\bibitem{Kogan}
S. Kogan, Phys. Rev. B {\bf 57},  9736 (1998)

%%%-----------------------------------------------------------------------------


\bibitem{Richards}
"Theory of one-dimensional hopping conductivity and diffusion",
P.M. Richards, Phys. Rev. B, {\bf 16}, 1393 (1977)

\bibitem{Thouless}
"Correlation effects in hopping conductivity",
K.S. Chase, D.J. Thouless, Phys. Rev. B {\bf 39} 9809 (1989)

\bibitem{Pitis91}
"Correlation corrections to the conductivity of one-dimensional disordered hopping models"
R. Pitis, P. Gartner, Phys. Rev. B {\bf 43} 11294 (1991)
%% 1D chains with two kinds of sites
%% close correlations <n_i n_{i+1}> considered
%% corrections to conductivity found

\bibitem{Pitis92}
"Occupancy-correlation corrections in hopping"
Phys. Rev. B {\bf 45} 7739 (1992)
P. Gartner, R. Pitis
%% Same systems, diffusion coefficient considered
%% approximations ??? (sophisticated formula)


\bibitem{PitisMonte93}
"Charge correlation factor of the random binary chain determined by Monte Carlo simulation"
R. Pitis, Phys. Rev. B {\bf 47} 15290 (1993)
%% 1D chain, two type of sites A and B, randomly arranged
%% studied: f = \sigma/ \sigma_{MF} MF=mean field; Monte-Carlo method
%% found f<1 (but f~1), qulitative agreement with previous studies

\bibitem{Shklovskii-cor}
E.I. Levin, V.L. Nguen, B.I. Shklovskii, Zh. Eksp. Theor.
Fiz. {\bf 82}, 1591 (1982);
Sov. Phys. JETP {\bf 55}  921 (1982)


\bibitem{Aleiner1}
O. Agam, I.L. Aleiner, Phys. Rev. B {\bf 89}, 224204 (2014)


\bibitem{pol-cor}
``Charge correlations in polaron hopping through molecules'',
B.B. Schmidt, M.H. Hettler, G. Schon,
Phys. Rev. B {\bf 82}, 155113 (2010)


\bibitem{AVS-cor}
Shumilin, Kabanov, Dedieu,
Phys. Rev. B {\bf 97}, 094201 (2018)

\bibitem{OMAR}
O. Mermer, G. Veeraraghavan, T. L. Francis, Y. Sheng, D. T. Nguyen, M. Wohlgenannt,
 A. Kohler,  M. K. Al-Suti, M. S. Khan, Phys. Rev. B {\bf 72}, 205202 (2005)

\bibitem{OMAR2} J. Kalinowski, M. Cocchi, D. Virgili, P. D. Marco, and V. Fattori,
Chem. Phys. Lett. {\bf 380}, 710 (2003).


\bibitem{OMAR3} V. N. Prigodin, J. D. Bergeson, D. M. Lincoln, and A. J. Epstein,
Synth. Met. {\bf 156}, 757 (2006).


\bibitem{Bobbert} P. A. Bobbert, T. D. Nguyen, F. W. A. van Oost,
B. Koopmans, and M. Wohlgenannt, Phys. Rev. Lett. {\bf 99}, 216801 (2007).

\bibitem{AVS-MF}
A.V. Shumilin, V.V. Kabanov, Phys. Rev. B {\bf 92}, 014206 (2015)



\bibitem{Bobbert-Res}
S.P. Kersten, S.C.J. Meskers, P.A. Bobbert, Phys. Rev. B {\bf 86}, 045210 (2012)
%%  used Thouless resistors in small part of the system
%%  to explain OMAR

\bibitem{dis-org}
A. Larabi, D. Bourbie,
J. Appl. Phys. {\bf 121}, 085502 (2017)

\bibitem{Aleiner2}
O. Agam, I.L. Aleiner, B. Spivak, Phys. Rev. B {\bf 89}, 100201(R) (2014)


\bibitem{Bogolubov}
Bogolubov, N. N. (1946): ``Problemy dinamicheskoii teorii v statisticheskoi
phisike'', (in Russian), Moscow

\bibitem{Landau10}
Landau, L. D. and Lifshitz, E. M. (1995): Physical Kinetics, Butter-
worth/Heinemann

\bibitem {Balescu} R. Balescu, Equilibriun and Nonequilibrium Statistical Mechan-
ics, A Wiley-Interscience Publications (Wiley, 1975)

\bibitem{sup} Supplementary materials. The archive with current distributions calculated in different
numerical samples with Miller-Abrahams, ``cor1'' and ``cor2'' approximations. The folders in the archive correspond to the samples.
The files inside one folder correspond to different approximations.

\end{thebibliography}
\end{document}